\documentclass[11pt,a4paper]{article}
\usepackage[latin1]{inputenc}
\usepackage[totalwidth=410pt, totalheight=595pt]{geometry}
\usepackage{cite}

\usepackage{amsmath}
\usepackage{amsfonts}
\usepackage{amssymb}

\numberwithin{equation}{section}

\newcommand {\beq} {\begin{equation}}
\newcommand {\eeq} {\end{equation}}

\newcommand {\ph}[1]{\phantom{#1}}

\newcommand {\sss} {\scriptscriptstyle}

\newcommand{\al}{\ensuremath{\alpha}}
\newcommand{\be}{\ensuremath{\beta}}
\newcommand{\ga}{\ensuremath{\gamma}}

\newcommand{\de}{\ensuremath{\delta}}

\newcommand{\eps}{\ensuremath{\epsilon}}

\newcommand{\la}{\ensuremath{\lambda}}

\newcommand{\om}{\ensuremath{\omega}}
\newcommand{\Om}{\ensuremath{\Omega}}

\newcommand{\alh}{\ensuremath{\hat{\alpha}}}

\newcommand{\ald}{\ensuremath{\dot{\alpha}}}
\newcommand{\bed}{\ensuremath{\dot{\beta}}}
\newcommand{\gad}{\ensuremath{\dot{\gamma}}}
\newcommand{\ded}{\ensuremath{\dot{\delta}}}
\newcommand{\mud}{\ensuremath{\dot{\mu}}}
\newcommand{\nud}{\ensuremath{\dot{\nu}}}
\newcommand{\rhod}{\ensuremath{\dot{\rho}}}
\newcommand{\sid}{\ensuremath{\dot{\sigma}}}
\newcommand{\alp}{\ensuremath{\acute{\alpha}}}
\newcommand{\bep}{\ensuremath{\acute{\beta}}}

\newcommand{\ad}{\ensuremath{\dot{a}}}
\newcommand{\bd}{\ensuremath{\dot{b}}}
\newcommand{\cd}{\ensuremath{\dot{c}}}
\newcommand{\ap}{\ensuremath{\acute{a}}}
\newcommand{\bp}{\ensuremath{\acute{b}}}

\newcommand{\kde}[2]{\de_{#1}^{\ph{#1}#2}}
\newcommand{\dia}[3]{{#1}_{#2}^{\ph{#2}#3}}

\newcommand{\mathHb}[1]{{\mathop{\kern0pt#1}\limits^{\,\sss
      \prime\prime}\vphantom{#1}}}

\newcommand{\com}[2]{\Big[ #1 , #2 \Big]}

\newcommand{\eqnlab}[1]{\label{eqn:#1}}

\newcommand{\eqnref}[1]{(\ref{eqn:#1})}

\newcommand{\Eqnref}[1]{Eq.~(\ref{eqn:#1})}

\newcommand{\Eqsref}[1]{Eqs.~(\ref{eqn:#1})}

\newcommand{\seclab}[1]{\label{sec:#1}}
\newcommand{\Secref}[1]{Section~\ref{sec:#1}}

\begin{document}

 \pagestyle{empty}

\begin{center}

\vspace*{2cm}

\noindent
{\LARGE\textsf{\textbf{BPS surface observables in \\[5mm] six-dimensional (2,0) theory}}}
\vskip 3truecm

{\large \textsf{\textbf{P\"ar Arvidsson}}} \\
\vskip 1truecm
{\it Department of Fundamental Physics\\ Chalmers University of
  Technology \\ SE-412 96 G\"{o}teborg,
  Sweden}\\[3mm] {\tt par.arvidsson@chalmers.se} \\
\end{center}
\vskip 1cm
\noindent{\bf Abstract:}
The supergroup $OSp(8^*|4)$, which is the superconformal group of $(2,0)$ theory in six dimensions, is broken to the subgroup $OSp(4|2) \times OSp(4|2)$ by demanding the invariance of a certain product in a superspace with eight bosonic and four fermionic dimensions. We show that this is consistent with the symmetry breaking induced by the presence of a flat two-dimensional BPS surface in the usual $(2,0)$ superspace, which has six bosonic and sixteen fermionic dimensions.

\newpage
\pagestyle{plain}


\section{Introduction}

One of the main motivations for the study of superconformal $(2,0)$ theories in six dimensions is their relation to maximally supersymmetric $\mathcal{N}=4$ Yang-Mills theory in four dimensions~\cite{Witten:1995,Verlinde:1995,Green:1996,Henningson:2000cs,Witten:2002}. This viewpoint provides an important insight into the $S$-duality of the four-dimensional theory~\cite{Montonen:1977,Goddard:1977,Osborn:1979}, since the electrically and magnetically charged particles are interpreted as different windings of self-dual strings on the compactification torus. $S$-duality then becomes a simple consequence of the $SL(2,\mathbb{Z})$ modular invariance.

Another intriguing connection between these theories is the following: In the four-dimensional $\mathcal{N}=4$ Yang-Mills theory, there are two important observables that have attracted much interest recently. These are the Wilson operator and its dual, the 't~Hooft operator, which both are associated with closed spatial curves. From the six-dimensional perspective, these should correspond to different windings of a single type of BPS \emph{surface} observable in $(2,0)$ theory~\cite{Ganor:1996,Maldacena:1998w,Berenstein:1998,Corrado:1999}. The connection between the Wilson and the 't~Hooft operators is also indicated by the results in Refs.~\cite{Henningson:2001cr,Henningson:2006}.

The study of BPS Wilson loops in $\mathcal{N}=4$ supersymmetric Yang-Mills theory is also motivated from the AdS/CFT correspondence~\cite{Maldacena:1998,Drukker:1999,Drukker:2000}. This viewpoint provides important information about string theory on $AdS_5 \times S^5$. Similarly, the study of BPS surface observables in $(2,0)$ theory might be related to quantities in $M$-theory on $AdS_7 \times S^4$.

The purpose of the present paper is to investigate in what way the superconformal group $OSp(8^*|4)$ is broken if a spatial two-dimensional flat surface is introduced in the $(2,0)$ superspace. This is inspired by the work in Ref.~\cite{Bianchi:2002}, where it is shown that a line in four-dimensional $\mathcal{N}=4$ Yang-Mills theory breaks the superconformal group $PSU(2,2|4)$ to the subgroup $OSp(2,2|4)$. We will use the results obtained in Refs.~\cite{Arvidsson:2006,Thesis} to simplify the treatment of the superconformal transformations, and we expect that the results obtained here will be useful in future studies of surface operators in $(2,0)$ theory.

The outline of this paper is as follows: \Secref{superc} reviews the superconformal group and the transformations therein, focusing on the possibility of a linear formulation in a superspace with eight bosonic and four fermionic dimensions. In \Secref{breaking}, we introduce a way of breaking the superconformal symmetry of the theory and show that the remaining unbroken symmetry leaves a two-dimensional surface invariant. Finally, \Secref{manifest} contains an attempt to describe this surface from an eight-dimensional perspective.

\section{The superconformal group}
\seclab{superc}

Consider the supergroup $OSp(8^*|4)$, which is the superconformal group relevant for $(2,0)$ theory in six dimensions~\cite{Kac:1977,Nahm:1978}. By definition, this group leaves the inner product
\beq
\eqnlab{inner_prod}
y \cdot z \equiv I^{\sss AB} y_{\sss A} z_{\sss B}
\eeq
invariant. In this relation, $y_{\sss A}$ and $z_{\sss B}$ denote coordinate vectors in a superspace with eight bosonic and four fermionic dimensions, while the graded symmetric tensor $I^{\sss AB}$ is the corresponding (inverse) metric. The superindices and the metric will be further explained below.

The generator of superconformal transformations is denoted by $J_{\sss AB}$, which is graded antisymmetric and obeys the (anti)commutation relations
\beq
\Big[ J_{\sss AB},J_{\sss CD} \Big\} = -\frac{1}{2} \Big( I_{\sss BC}
J_{\sss AD} - (-1)^{\sss AB} I_{\sss AC} J_{\sss BD} -
(-1)^{\sss CD} I_{\sss BD} J_{\sss AC} +
(-1)^{\sss AB+CD} I_{\sss AD} J_{\sss BC} \Big),
\eqnlab{superalgebra}
\eeq
where the bracket in the left hand side is an anticommutator if both its entries are fermionic, otherwise it is a commutator. A factor $(-1)^{\sss A}$ is positive if $A$ is a bosonic index, and negative if it is fermionic.

The corresponding coordinate transformation is given by
\beq
\de y_{\sss A} = - \pi^{\sss CD} y_{\sss C} I_{\sss DA},
\eqnlab{coord_transf}
\eeq
where the graded antisymmetric quantity $\pi^{\sss CD}$ contains the infinitesimal parameters. It is easily verified that a transformation of this form indeed leaves the inner product in \Eqnref{inner_prod} invariant.

It is illustrative to decompose the quantities introduced above into more familiar ones~\cite{Arvidsson:2005,Arvidsson:2006,Thesis}. The coordinate vector may be written as $y_{\sss A}=(y_{\alh},y^a)=(y_\al,y^\al,y^a)$. In this expression, the bosonic index $\alh=(1,\ldots,8)$ is a chiral $SO(6,2)$ spinor index, which may be further decomposed into one chiral $SO(5,1)$ spinor index $\al=(1,\ldots,4)$ (subscript) and one anti-chiral $SO(5,1)$ spinor index $\al=(1,\ldots,4)$ (superscript). Finally, the fermionic index $a=(1,\ldots,4)$ is an $SO(5)$ spinor index.

In this notation, the superconformal generators become~\cite{Claus:1998,Arvidsson:2005}
\beq
  J_{\sss AB} =
  \left( \begin{array}{ccc}
  \frac{1}{2} P_{\al\be} &  \frac{1}{2} \dia{M}{\al}{\be} + \frac{1}{4} \kde{\al}{\be} D & \frac{i}{2\sqrt{2}} Q^b_\al \\
  - \frac{1}{2} \dia{M}{\be}{\al} - \frac{1}{4} \kde{\be}{\al} D & - \frac{1}{2} K^{\al\be} & \frac{i}{2\sqrt{2}} \Om^{bc} S^\al_c \\
  - \frac{i}{2\sqrt{2}} Q^a_\be & - \frac{i}{2\sqrt{2}} \Om^{ac} S^\be_c & i U^{ab}
  \end{array} \right),
  \eqnlab{J_AB}
\eeq
while the superspace metric is written as
\beq
  I_{\sss AB} =
  \left( \begin{array}{ccc}
  0 &  \kde{\al}{\be} & 0 \\
  \de^\al_{\ph{\al}\be} & 0 & 0 \\
  0 & 0 & i \Om^{ab}
  \end{array} \right).
  \eqnlab{superspacemetric}
\eeq
Together with \Eqnref{superalgebra}, these definitions reproduce all commutation relations of the six-dimensional superconformal algebra with conventions as in Ref.~\cite{Arvidsson:2005}. We have also introduced the $SO(5)$ invariant antisymmetric tensor $\Om^{ab}$ in the purely fermionic piece of the superspace metric.

In the inner product defined in \Eqnref{inner_prod}, the inverse superspace metric appears, i.e., the metric with superscript indices. This is written as
\beq
  I^{\sss AB} =
  \left( \begin{array}{ccc}
  0 &  \de^\al_{\ph{\al}\be} & 0 \\
  \kde{\al}{\be} & 0 & 0 \\
  0 & 0 & -i \Om_{ab}
  \end{array} \right)
  \eqnlab{superspacemetric_upper}
\eeq
in this basis, which makes the relation
\beq
I_{\sss AB} I^{\sss BC} = \de^{\sss C}_{\sss A}
\eeq
valid (which is essential if we want to raise and lower indices). This requires that $\Om_{ab} \Om^{bc} = \kde{a}{c}$.

\section{Breaking the superconformal symmetry}
\seclab{breaking}

Consider the subgroup $H \subset OSp(8^*|4)$ which leaves the product
\beq
y \circ z \equiv \hat{I}^{\sss AB} y_{\sss A} z_{\sss B}
\eqnlab{brokenproduct}
\eeq
invariant. In this expression, we demand the matrix $\hat{I}^{\sss AB}$ to satisfy
\beq
\begin{aligned}
I_{\sss AB} \hat{I}^{\sss AB} &= 0 \eqnlab{tracehat} \\
I_{\sss BC} \hat{I}^{\sss AB} \hat{I}^{\sss CD} &= I^{\sss AD}.
\end{aligned}
\eeq
Explicitly, we choose our basis such that
\beq
  \hat{I}^{\sss AB} =
  \left( \begin{array}{cccccc}
  0 & 0 & \de^{\ald}_{\ph{\ald}\bed} & 0 & 0 & 0 \\
  0 & 0 & 0 & - \de^{\alp}_{\ph{\alp}\bep} & 0 & 0 \\
  \kde{\ald}{\bed} & 0 & 0 & 0 & 0 & 0 \\
  0 & - \kde{\alp}{\bep} & 0 & 0 & 0 & 0 \\
  0 & 0 & 0 & 0 & \eps_{\ad \bd} & 0 \\
  0 & 0 & 0 & 0 & 0 & - \eps_{\ap \bp}
  \end{array} \right),
  \eqnlab{brokenmetric}
\eeq
where all indices in the matrix are fundamental $SU(2)$ indices, taking the values $1$ and $2$. In this basis, the $SO(5)$ invariant tensor is written as
\beq
  \Om_{ab} =
  \left( \begin{array}{cc}
  i \eps_{\ad \bd} & 0 \\
  0 & i \eps_{\ap \bp}
  \end{array} \right),
\eeq
where the antisymmetric $SU(2)$ invariant tensor $\eps_{\ad\bd}$ is defined such that $\eps_{12}=1$. This means that we have decomposed the $SO(5,1)$ spinor indices according to $\al=(\ald,\alp)$, but also the $SO(5)$ spinor index as $a=(\ad,\ap)$. In total, this leaves us with six different $SU(2)$ indices. Note that the indices denoted by Greek letters, originating from the bosonic piece of the superindex, may not be raised or lowered; superscript and subscript indices indicate different representations.

Comparing the expression for $\hat{I}^{\sss AB}$ in \Eqnref{brokenmetric} with the inverse superspace metric $I^{\sss AB}$ in \Eqnref{superspacemetric_upper}, we see that they differ only by some signs. The inspiration for this may be found in Ref.~\cite{AFH:2003free}, where the $R$-symmetry group $SO(5)$ is spontaneously broken to $SO(4) \simeq SU(2) \times SU(2)$ by selecting a specific unit vector $\hat{\Phi}^{ab}$ as
\beq
  \hat{\Phi}^{ab} =
  \left( \begin{array}{cc}
  i \eps^{\ad \bd} & 0 \\
  0 & - i \eps^{\ap \bp}
  \end{array} \right).
\eeq
The quantity in \Eqnref{brokenmetric} generalizes this idea to the entire superconformal space. Specifically, we note that \Eqnref{tracehat} is analogous to the relation $\Om_{ab} \hat{\Phi}^{ab}=0$ from Ref.~\cite{AFH:2003free}, which makes $\hat{\Phi}^{ab}$ an $SO(5)$ vector.

To investigate the properties of the subgroup $H$, let us see which $OSp(8^*|4)$ generators leave the product in \Eqnref{brokenproduct} invariant. By applying the transformation in \Eqnref{coord_transf}, we find that the product is invariant if the generator $J_{\sss AB}$ is of the form
\beq
\hat{J}_{\sss AB} = \left( \begin{array}{cccccc}
  J_{\ald\bed} & 0 & \dia{J}{\ald}{\bed} & 0 & \dia{J}{\ald}{\bd} & 0 \\
  0 & J_{\alp\bep} & 0 & \dia{J}{\alp}{\bep} & 0 & \dia{J}{\alp}{\bp} \\
  J^{\ald}_{\ph{\ald}\bed} & 0 & J^{\ald\bed} & 0 & J^{\ald \bd} & 0 \\
  0 & J^{\alp}_{\ph{\alp}\bep} & 0 & J^{\alp\bep} & 0 & J^{\alp \bp} \\
  J^{\ad}_{\ph{\ald}\bed} & 0 & J^{\ad \bed} & 0 & J^{\ad \bd} & 0 \\
  0 & J^{\ap}_{\ph{\alp}\bep} & 0 & J^{\ap \bep} & 0 & J^{\ap \bp}
  \end{array} \right).
  \eqnlab{brokenJ}
\eeq
By rearranging the indices, this matrix is easily brought to a block-diagonal form, indicating that the subgroup $H$ is a product of two identical groups; one associated with dotted and one with primed indices. These groups each have nine bosonic and eight fermionic generators, to be compared with the 38 bosonic and 32 fermionic generators of $OSp(8^*|4)$. Thus, half the supersymmetries and half the special supersymmetries are left unbroken.

Next, consider how $H$ acts on the $(2,0)$ superspace with six bosonic coordinates, denoted by $x^{\al\be}=-x^{\be\al}$, and sixteen fermionic coordinates, $\theta^\al_a$. The coordinate transformations induced by the full $OSp(8^*|4)$ supergroup are given by~\cite{Arvidsson:2005,Park:1998}
\begin{align}
\eqnlab{transf_superx}
\begin{split}
\de x^{\al \be} &= a^{\al \be} - \dia{\om}{\ga}{[\al} x^{\be] \ga} +
\la x^{\al \be} + 4 c_{\ga \de} x^{\ga\al} x^{\be\de} - i \Om^{ab} \eta^{[\al}_{a\ph{b}} \theta^{\be]}_b - {} \\
& \quad - c_{\ga \de} \theta^\ga \cdot \theta^{[\al}
\theta^{\be]} \cdot \theta^{\de} -i \rho^c_{\ga} \theta^{[\al}_c
    \left( 2 x^{\be]\ga} - i \theta^{\be]} \cdot \theta^{\ga} \right)
\end{split} \\
\eqnlab{transf_superth}
\begin{split}
\de \theta^{\al}_a &= (\dia{\om}{\ga}{\al} - 4 c_{\ga \de} x^{\al \de}
- 2i c_{\ga \de} \theta^{\al} \cdot \theta^{\de} + 2i
\rho^c_{\ga} \theta^{\al}_c) \theta^{\ga}_a + \frac{1}{2} \la \theta^{\al}_a + {} \\
& \quad + \eta^{\al}_a - \Om_{ac} \rho^c_{\ga} \left( 2 x^{\ga \al} - i \theta^{\ga} \cdot \theta^{\al} \right) + v_{ac} \Om^{cd} \theta^{\al}_d,
\end{split}
\end{align}
where the conventions for the various parameters may be found in Ref.~\cite{Arvidsson:2005}. If we split the indices on the coordinates and restrict ourselves to the transformations contained in $H$, we find that a configuration where
\begin{equation}
\begin{split}
x^{\ald\bep} &= 0 \\
\theta^{\ald}_{\bp} &= 0 \\
\theta^{\alp}_{\bd} &= 0
\end{split}
\eqnlab{surface}
\end{equation}
is left invariant. This corresponds naturally to a flat two-dimensional surface embedded in the six-dimensional space-time, which in addition breaks half the supersymmetry of the theory and also breaks the $R$-symmetry group. The remaining bosonic symmetry transformations, contained in the supergroup $H$, are interpreted as conformal transformations on the surface and rotations in the transverse space, together with the $R$-symmetry rotations in the space transverse to $\hat{\Phi}^{ab}$.

It is interesting to note that the surface specified in \Eqnref{surface} is necessarily associated to a direction in $R$-symmetry space. This means that the presence of the surface automatically breaks the $R$-symmetry group from $SO(5)$ to $SO(4)$. This phenomenon is a consequence of the breaking of supersymmetry.

Let us investigate the properties of the supergroup $H$: The bosonic subgroup mentioned above is $SO(3,1)\times SO(3,1) \times SO(4)$, provided that the surface is spatial. This means quite naturally that the bosonic subgroup of each of the two factors in $H$ is $SO(3,1)\times SU(2)$. The most plausible supergroup with this bosonic subgroup is $OSp(3,1|2)$; the next step is to fit the generators contained in $\hat{J}_{\sss AB}$ into this structure.

Leaving questions concerning signature aside, the supergroup $OSp(4|2)$ is defined through the commutation relations
\begin{equation}
\begin{aligned}
\com{A_{\ald \bed}}{R^{\ad}_{\gad,\mud}} &= \eps^{\ph{\ad}}_{\gad(\ald} R^{\ad}_{\bed),\mud} \qquad
& \com{A_{\ald \bed}}{A_{\gad \ded}} &= \eps_{\gad(\ald} A_{\bed)\ded} + \eps_{\ded(\ald} A_{\bed)\gad} \\
\com{B_{\mud \nud}}{R^{\ad}_{\ald,\rhod}} &= \eps^{\ph{\ad}}_{\rhod(\mud} R^{\ad}_{\ald,\nud)}
& \com{B_{\mud \nud}}{B_{\rhod \sid}} &= \eps_{\rhod(\mud} B_{\nud)\sid} + \eps_{\sid(\mud} B_{\nud)\rhod} \\
\com{C^{\ad \bd}}{R^{\cd}_{\ald,\mud}} &= \eps^{\cd (\ad} R^{\bd)}_{\ald,\mud}
& \com{C^{\ad \bd}}{C^{\cd \dot{d}}} &= \eps^{\cd(\ad}_{\ph{\ald}} C^{\bd)\dot{d}} + \eps^{\dot{d}(\ad} C^{\bd)\cd},
\end{aligned}
\end{equation}
and the anti-commutation relation
\begin{equation}
\Big\{ R^{\ad}_{\ald,\mud}, R^{\bd}_{\bed,\nud} \Big\} = \eps^{\ad \bd} \eps_{\mud \nud} A_{\ald \bed} + \eps^{\ad \bd} \eps_{\ald \bed} B_{\mud \nud} - 2 \eps_{\ald\bed} \eps_{\mud\nud} C^{\ad\bd},
\eqnlab{anticom}
\end{equation}
where we note three three-component bosonic generators (denoted by $A_{\ald\bed}$, $B_{\mud\nud}$ and $C^{\ad \bd}$, all symmetric) and one eight-component fermionic generator $R^{\ad}_{\ald,\mud}$. The bosonic generators appear with different weights in the right-hand side of the anticommutator in \Eqnref{anticom}. Similar relations hold for the version with primed indices.

Let us see if we may fit these generators into the structure for $\hat{J}_{\sss AB}$ in \Eqnref{brokenJ}. This procedure ultimately fixes the weights in the right-hand side of \Eqnref{anticom} to these particular values. Define
\beq
\begin{aligned}
J_{\ald}^{\ph{\ald}\ad} &= \frac{1}{2} \left( R^{\ad}_{\ald,1} + R^{\ad}_{\ald,2} \right) \\
J^{\ald \ad} &= - \frac{1}{4} \eps^{\ald \gad} \left( R^{\ad}_{\gad,1} - R^{\ad}_{\gad,2} \right),
\end{aligned}
\eeq
where we have let the $\mud$-index take specific values. By requiring these generators to satisfy \Eqnref{superalgebra}, we find that we need to take
\beq
\begin{aligned}
J_{\ald \bed} &= - \frac{1}{2} \eps_{\ald\bed} \left( B_{11} + 2 B_{12} + B_{22} \right) \\
J^{\ald \bed} &= - \frac{1}{8} \eps^{\ald\bed} \left( B_{11} - 2 B_{12} + B_{22} \right) \\
J_{\ald}^{\ph{\ald}\bed} &= - \frac{1}{2} \eps^{\bed \gad} A_{\ald\gad} + \frac{1}{4} \kde{\ald}{\bed} \left(B_{11} - B_{22} \right) \\
J^{\ad \bd} &= - C^{\ad \bd}.
\end{aligned}
\eeq
In this way, we have constructed an explicit isomorphism between the unbroken subgroup $H \subset OSp(8^*|4)$ and the supergroup $OSp(4|2) \times OSp(4|2)$.

So far, we have considered a flat infinitely extended two-dimensional surface. This is in fact a special case; in general we should consider closed surfaces, such as two-spheres. The moduli space of possible such surfaces is parametrized by the coset $OSp(8^*|4) / [OSp(4|2) \times OSp(4|2)]$, and has 20 bosonic and 16 fermionic dimensions.

\section{The superconformal space}
\seclab{manifest}

The purpose of this section is to investigate if there is another way, based on an eight-dimensional formulation, of showing that the product in \Eqnref{brokenproduct} is associated with a surface in six dimensions.

In Ref.~\cite{Arvidsson:2006}, the connection between the superconformal space (with eight bosonic dimensions) and the $(2,0)$ superspace (with six bosonic dimensions) was accomplished by requiring the fields to live on a projective supercone, defined by
\beq
I^{\sss AB} y_{\sss A} y_{\sss B}=0.
\eqnlab{supercone}
\eeq
An explicit solution to this condition is given by
\beq
\eqnlab{y-x}
\begin{aligned}
y^a &= \sqrt{2} \Om^{ab} \theta_b^\be y_\be \\
y^\al &= \left( 2 x^{\al\be} - i \Om^{ab} \theta^\al_a \theta^\be_b \right) y_\be.
\end{aligned}
\eeq
The consistency of these equations implies that $x^{\al\be}$ and $\theta^\al_a$ must transform according to \Eqsref{transf_superx} and \eqnref{transf_superth}, respectively, if the $y_{\sss A}$ coordinates are transformed as in \Eqnref{coord_transf}. This indicates that we may identify these quantities with the usual coordinates in $(2,0)$ superspace.

Let us define a new constraint, similar to \Eqnref{supercone}, but this time based on the product in \Eqnref{brokenproduct}. The most natural such condition is
\beq
\hat{I}^{\sss AB} y_{\sss A} y_{\sss B}=0.
\eqnlab{condition}
\eeq
If we combine this condition with the supercone constraint \eqnref{supercone}, we find two separate equations:
\beq
\begin{aligned}
2 y^{\ald} y_{\ald} + \eps_{\ad \bd} y^{\ad} y^{\bd} &=0 \\
2 y^{\alp} y_{\alp} + \eps_{\ap \bp} y^{\ap} y^{\bp} &=0.
\end{aligned}
\eeq
In analogy with \Eqnref{y-x}, we write an explicit solution to these equations as
\beq
\eqnlab{y-x_2}
\begin{aligned}
y^{\ad} &= \sqrt{2} i \eps^{\ad \bd} \theta_{\bd}^{\bed} y_{\bed} \\
y^{\ald} &= \left( 2 x^{\ald \bed} + \eps^{\ad \bd} \theta^{\ald}_{\ad} \theta^{\bed}_{\bd} \right) y_{\bed},
\end{aligned}
\eeq
together with similar equations with primed indices. We see immediately that this defines a surface with two bosonic dimensions, which also is described by eight fermionic variables. This agrees with the results found in the preceding section and suggests that the surface in $(2,0)$ superspace may be regarded as the intersection of the supercone \eqnref{supercone} and the hyper-surface \eqnref{condition} in the superconformal space.

\vspace{7mm}
\noindent
\textbf{Acknowledgments:}
The author would like to thank M{\aa}ns Henningson for stimulating and encouraging discussions.
\bibliographystyle{utphysmod3b}
\addcontentsline{toc}{section}{References}
\bibliography{biblio}

\providecommand{\href}[2]{#2}\begingroup\raggedright\begin{thebibliography}{10}

\bibitem{Witten:1995}
E.~Witten,  {\em Some comments on string dynamics},
\href{http://arXiv.org/abs/hep-th/9507121}{{\tt hep-th/9507121}}.

\bibitem{Verlinde:1995}
E.~Verlinde,  {\em Global aspects of electric-magnetic duality}, Nucl. Phys.
  {\bf B455} (1995) 211--228
[\href{http://www.arXiv.org/abs/hep-th/9506011}{{\tt hep-th/9506011}}].

\bibitem{Green:1996}
M.~B. Green, N.~D. Lambert, G.~Papadopoulos and P.~K. Townsend,  {\em Dyonic
  p-branes from self-dual (p+1)-branes}, Phys. Lett. {\bf B384} (1996) 86--92
[\href{http://www.arXiv.org/abs/hep-th/9605146}{{\tt hep-th/9605146}}].

\bibitem{Henningson:2000cs}
M.~Henningson,  {\em A class of six-dimensional conformal field theories},
  Phys. Rev. Lett. {\bf 85} (2000) 5280
[\href{http://arXiv.org/abs/hep-th/0006231}{{\tt hep-th/0006231}}].

\bibitem{Witten:2002}
E.~Witten, {\em Conformal field theory in four and six dimensions}.
\newblock Prepared for Symposium on Topology, Geometry and Quantum Field Theory
  (Segalfest), Oxford, England, United Kingdom, 24-29 Jun 2002.

\bibitem{Montonen:1977}
C.~Montonen and D.~I. Olive,  {\em Magnetic monopoles as gauge particles?},
  Phys. Lett. {\bf B72} (1977)
117.

\bibitem{Goddard:1977}
P.~Goddard, J.~Nuyts and D.~I. Olive,  {\em Gauge theories and magnetic
  charge}, Nucl. Phys. {\bf B125} (1977)
1.

\bibitem{Osborn:1979}
H.~Osborn,  {\em Topological charges for {N}=4 supersymmetric gauge theories
  and monopoles of spin 1}, Phys. Lett. {\bf B83} (1979)
321.

\bibitem{Ganor:1996}
O.~J. Ganor,  {\em Six-dimensional tensionless strings in the large {N} limit},
  Nucl. Phys. {\bf B489} (1997) 95--121
[\href{http://www.arXiv.org/abs/hep-th/9605201}{{\tt hep-th/9605201}}].

\bibitem{Maldacena:1998w}
J.~Maldacena,  {\em Wilson loops in large {N} field theories}, Phys. Rev. Lett.
  {\bf 80} (1998) 4859--4862
[\href{http://www.arXiv.org/abs/hep-th/9803002}{{\tt hep-th/9803002}}].

\bibitem{Berenstein:1998}
D.~Berenstein, R.~Corrado, W.~Fischler and J.~Maldacena,  {\em The operator
  product expansion for {W}ilson loops and surfaces in the large {N} limit},
  Phys. Rev. {\bf D59} (1999) 105023
[\href{http://www.arXiv.org/abs/hep-th/9809188}{{\tt hep-th/9809188}}].

\bibitem{Corrado:1999}
R.~Corrado, B.~Florea and R.~McNees,  {\em Correlation functions of operators
  and {W}ilson surfaces in the d = 6, (0,2) theory in the large {N} limit},
  Phys. Rev. {\bf D60} (1999) 085011
[\href{http://www.arXiv.org/abs/hep-th/9902153}{{\tt hep-th/9902153}}].

\bibitem{Henningson:2001cr}
M.~Henningson,  {\em Commutation relations for surface operators in
  six-dimensional (2,0) theory}, J. High Energy Phys. {\bf 03} (2001) 011
[\href{http://arXiv.org/abs/hep-th/0012070}{{\tt hep-th/0012070}}].

\bibitem{Henningson:2006}
M.~Henningson,  {\em Wilson-'t {H}ooft operators and the theta angle}, JHEP
  {\bf 05} (2006) 065
[\href{http://www.arXiv.org/abs/hep-th/0603188}{{\tt hep-th/0603188}}].

\bibitem{Maldacena:1998}
J.~M. Maldacena,  {\em The large {N} limit of superconformal field theories and
  supergravity}, Adv. Theor. Math. Phys. {\bf 2} (1998) 231--252
[\href{http://www.arXiv.org/abs/hep-th/9711200}{{\tt hep-th/9711200}}].

\bibitem{Drukker:1999}
N.~Drukker, D.~J. Gross and H.~Ooguri,  {\em Wilson loops and minimal
  surfaces}, Phys. Rev. {\bf D60} (1999) 125006
[\href{http://www.arXiv.org/abs/hep-th/9904191}{{\tt hep-th/9904191}}].

\bibitem{Drukker:2000}
N.~Drukker and D.~J. Gross,  {\em An exact prediction of {$\mathcal{N}=4$}
  supersymmetric {Y}ang-{M}ills theory for string theory}, J. Math. Phys. {\bf
  42} (2001) 2896--2914
[\href{http://www.arXiv.org/abs/hep-th/0010274}{{\tt hep-th/0010274}}].

\bibitem{Bianchi:2002}
M.~Bianchi, M.~B. Green and S.~Kovacs,  {\em Instanton corrections to circular
  {W}ilson loops in {$\mathcal{N}=4$} supersymmetric {Y}ang-{M}ills}, J. High
  Energy Phys. {\bf 04} (2002) 040
[\href{http://www.arXiv.org/abs/hep-th/0202003}{{\tt hep-th/0202003}}].

\bibitem{Arvidsson:2006}
P.~Arvidsson,  {\em Manifest superconformal covariance in six-dimensional (2,0)
  theory}, J. High Energy Phys. {\bf 03} (2006) 076
[\href{http://www.arXiv.org/abs/hep-th/0602193}{{\tt hep-th/0602193}}].

\bibitem{Thesis}
P.~Arvidsson, {\em Superconformal theories in six dimensions}.
\newblock PhD thesis, Chalmers University of Technology, 2006,
\href{http://www.arXiv.org/abs/hep-th/0608014}{{\tt hep-th/0608014}}.
\newblock

\bibitem{Kac:1977}
V.~G. Kac,  {\em A sketch of {L}ie superalgebra theory}, Commun. Math. Phys.
  {\bf 53} (1977)
31--64.

\bibitem{Nahm:1978}
W.~Nahm,  {\em Supersymmetries and their representations}, Nucl. Phys. {\bf
  B135} (1978)
149.

\bibitem{Arvidsson:2005}
P.~Arvidsson,  {\em Superconformal symmetry in the interacting theory of (2,0)
  tensor multiplets and self-dual strings}, J. Math. Phys. {\bf 47} (2006)
  042301
[\href{http://www.arXiv.org/abs/hep-th/0505197}{{\tt hep-th/0505197}}].

\bibitem{Claus:1998}
P.~Claus, R.~Kallosh and A.~Van~Proeyen,  {\em M5-brane and superconformal
  (0,2) tensor multiplet in 6 dimensions}, Nucl. Phys. {\bf B518} (1998)
  117--150
[\href{http://www.arXiv.org/abs/hep-th/9711161}{{\tt hep-th/9711161}}].

\bibitem{AFH:2003free}
P.~Arvidsson, E.~Flink and M.~Henningson,  {\em Free tensor multiplets and
  strings in spontaneously broken six-dimensional (2,0) theory}, J. High Energy
  Phys. {\bf 06} (2003) 039
[\href{http://www.arXiv.org/abs/hep-th/0306145}{{\tt hep-th/0306145}}].

\bibitem{Park:1998}
J.-H. Park,  {\em Superconformal symmetry in six-dimensions and its reduction
  to four}, Nucl. Phys. {\bf B539} (1999) 599--642
[\href{http://www.arXiv.org/abs/hep-th/9807186}{{\tt hep-th/9807186}}].

\end{thebibliography}\endgroup

\end{document}